\documentclass[debug, preprint, twocolumn]{rmaa}
\usepackage{natbib}
\usepackage{graphicx} 
\usepackage{amsmath}
\usepackage{hyperref}
\usepackage{longtable}

\usepackage{paralist}

\usepackage{psfrag,color}

\usepackage[latin1]{inputenc}

{\catcode`\$=\active
}%

\newcommand\Moon{\astro{\char10}}
\newcommand\Aries{\astro{\char11}}%
\newcommand\Libra{\astro{\char17}}%
\newcommand{\Cloudy}{\textsc{Cloudy}}

\font\manual=manfnt at 7pt \def\dbend{\hbox{\raise0.9ex\hbox{\manual\char127\hspace{0.6em}}}}

\providecommand{\e}[1]{\ensuremath{\times 10^{#1}}}

\newcounter{INTERNALionstage}

\def\gtsim{\mathrel{\hbox{\rlap{\hbox{\lower4pt\hbox{$\sim$}}}\hbox{$>$}}}}
\def\lesssim{\mathrel{\hbox{\rlap{\hbox{\lower4pt\hbox{$\sim$}}}\hbox{$<$}}}}

\def\A{{\rm\thinspace \AA}}

\def\g{{\rm\thinspace g}}

\def\K{{\rm\thinspace K}}

\def\m{{\rm\thinspace m}}

\def\s{{\rm\thinspace s}}

\def\W{{\rm\thinspace W}}

\def\h0{\mbox{{\rm H}$^0$}}
\DeclareMathAlphabet{\vib}{OML}{cmm}{m}{it}

\usepackage{common/aas_macros}

\title{Recent update of gas-phase chemical reactions and molecular lines of T\lowercase{i}O in \textsc{Cloudy}}

\author{Gargi Shaw\altaffilmark{1},
Gary J. Ferland\altaffilmark{2},
Phillip Stancil \altaffilmark{3},
Ryan Porter \altaffilmark{4}
}

\altaffiltext{1}{Department of Astronomy and Astrophysics, Tata Institute of Fundamental Research, Mumbai 400005, India}
\altaffiltext{2}{Physics \& Astronomy, University of Kentucky, Lexington, Kentucky, USA}
\altaffiltext{3}{Physics \& Astronomy, University of Georgia, Georgia, USA}
\altaffiltext{4}{Stellar Science, 6565 Americas Pkwy 925, Albuquerque, NM 87110, USA}

\fulladdresses{
\item G.{} Shaw, Department of Astronomy and Astrophysics, Tata Institute of Fundamental Research,  
Mumbai 400005, India (gargishaw@gmail.com).

 \item G.{} Ferland, Physics \& Astronomy, University of Kentucky, Lexington KY 40506, USA  (gary@uky.edu).

 \item P.{} Stancil, Department of Physics and Astronomy and Center for
Simulational Physics, University of Georgia, USA.

\item R.{} Porter, Stellar Science, 6565 Americas Pkwy 925, Albuquerque, NM 87110, USA
 
}
\shortauthor{Shaw et al.}
\shorttitle{Gas-phase T\lowercase{i}O}

\listofauthors{G. Shaw et al}

\indexauthor{Ferland, G.J.}
\indexauthor{Phillip Stancil}
\indexauthor{Ryan Porter}

\abstract{
We present our current update on the gas-phase chemical reactions and spectral 
lines of TiO in the spectral synthesis code \textsc{Cloudy}. 
For this purpose, we have added 229 Ti-related reactions in the chemical network. 
In addition, we consider 230 fine-structure energy levels, the corresponding 223 
radiative transitions, and 444 
collisional transitions with ortho and para H$_2$
and predict 66 TiO lines.
We perform spectroscopic simulations of TiO emission from the 
circumstellar region of the oxygen-rich red supergiant VY Canis Majoris to validate 
our update. Our model reproduces
the observed TiO column density. 

This update is helpful in modeling dust-free astrophysical environments where Ti is in the gas phase and TiO can form.
}
\resumen{
Presentamos nuestra actualizacion actual sobre las reacciones quimicas 
en fase gaseosa y espectrales lineas de TiO en el codigo de sintesis 
espectral \textsc{Cloudy}. 
Para ello, hemos anadido 229 reacciones relacionadas con 
el Ti en la red quimica. 
Ademas, consideramos 230 niveles 
de energia de estructura fina, los correspondientes 223 
transiciones radiativas, y 444 transiciones colisionales 
con orto y para H$_2$ y predecir 66 lineas de TiO. 
Realizamos simulaciones espectroscopicas de la emision de 
TiO desde el region circunestelar de la supergigante roja 
rica en oxigeno VY Canis Majoris para validar nuestra 
actualizacion. Nuestro modelo reproduce la densidad de la 
columna de TiO observada. Esta actualizacion es util para 
modelar entornos astrofisicos libres de polvo donde el 
Ti esta en fase gaseosa y se puede formar TiO.
}

\addkeyword{Astronomy software}
\addkeyword{Computational methods}
\addkeyword{Galaxy}
\addkeyword{Molecular data}

\begin{document}

\maketitle

\section{Introduction}
The spectroscopic simulation
code \textsc{Cloudy} simulates the conditions in
a non-equilibrium astrophysical plasma
and predicts the resulting spectrum and column densities of various ions, neutrals, 
and molecules\citep{{2023RMxAA..59..327C},{2023RNAAS...7..246G},{2017RMxAA..53..385F},{2013RMxAA..49..137F}}. 
Our aim is to predict more molecular lines with better precision. Hence, 
we regularly update our chemical network and incorporate improved chemical reactions
and internal structures for our existing chemistry whenever new or better data are available \citep{{2017ApJ...843..149S},{2020RNAAS...4...78S},{2023RNAAS...7...45S}, 2022ApJ...934...53S}
and add new molecules as the astrophysical need arises.
One such molecule is TiO, a possible precursor of inorganic dust \citep{2020ApJ...904..110D}. 

TiO is the dominant source of opacity in the atmosphere of cool stars \citep{Lodders_2002}
and is observed in the atmosphere of giant M stars \citep{1994A&A...284..179J, 2013ApJS..209...38K}. 
\citet{2013A&A...551A.113K} observed pure rotational lines 
of TiO in the circumstellar envelope of the red supergiant VY Canis Majoris (VY CMa). 
The observed column density of TiO in VY CMa is (6.7 $\pm$ 0.8) 
$\times$ 10$^{15}$ cm$^{-2}$. They also observed TiO$_2$. 
In an oxygen-rich circumstellar envelope, nearly all carbon is
locked into CO, and hence, TiO forms and can act as an important seed for 
inorganic dust formation \citep{1998FaDi..109..303G}. 
\citet {2013ApJS..209...38K} showed that for VY CMa, the dust and TiO emission come 
from different regions.
Their formation regions are separated by about 0$''$.15 corresponding to 2.7$\times$10$^{13}$cm assuming the 
distance of VY CMa is 1.20 kpc \citep{2012ApJ...744...23Z}.
This confirms that TiO is present in the dust-free region of VY CMa.

 The solar abundance of Ti is 8.91$\times$10$^{-8}$ \citet{2010Ap&SS.328..179G}. However, TiO is not observed in the ISM due to the high depletion of Ti \citep{1997ApJS..112..507W}.
In dust-free environments, Ti is in the gas phase and TiO can form.

Here, we aim to include the gas-phase chemical reactions and spectral lines of TiO 
in the spectral synthesis code Cloudy \citep{2023RMxAA..59..327C, 2023RNAAS...7..246G} which will be useful in modeling dust-free astrophysical environments where TiO is in the gas phase. These updates will be part of the upcoming 2024 release of Cloudy.

\section{Calculations and results}
\subsection{Updated T\lowercase{i}-chemistry in \Cloudy}
There is a scarcity of reaction rates for Ti-chemistry due to its very low abundance in the ISM. 
It is not included in the UMIST Database for Astrochemistry (UDfA) \citep{2013A&A...550A..36M} or in Kinetic Database for Astrochemistry (KIDA) \citep{2012ApJS..199...21W}. We incorporate available 
reactions\citep{{2021ApJ...923..264T},{1980AJ.....85.1382C},{2020A&A...641A..87R}}, but these are small in number. 

We adopted Si-based chemistry as a proxy for the missing Ti chemistry.
Both Ti and Si are important refractory elements \citep{2006A&A...449..723G}. 
In addition, 
Ti-oxides and Si-oxides are precursors of 
inorganic dust, and Si-chemistry is well-studied and documented.
Hence, for 
the rest of the Ti-related reactions, we copied analogous reactions involving Si from 
UDfA. All the reactions in our new network are given in the appendix. 


A large number of models covering diverse astrophysical environments are publicly 
available with the \textsc{cloudy} download under the directory \texttt{tsuite}.
We compare tsuite model predictions every day to monitor the changes in species
abundances and line intensities as a result of changes in the source
code and atomic and molecular data. The Cloudy test suite is extensive and 
includes simulations
at very low and very high temperatures, CMB to 10$^{10}$ K. 
The chemistry must work for all conditions.
Often we encounter problems with the extrapolation of simple
fits to the temperature range we need for chemical rate coefficients. 
To overcome this problem, we apply a 
temperature cap, T$_{cap}$, for $\beta$ $>$ 0 (See appendix). For T $>$ T$_{cap}$, 
the rate coefficients
retain the same values as at T$_{cap}$. Though ad-hoc, we choose 
T$_{cap}$=2500 K \citep{Shaw_2023}. 
The chemical network is very sensitive to small 
changes in rate coefficients. Hence, for any chemical update to pass through all the 
tsuite models simultaneously is hard and time-consuming. 
We critically examine each update if it results in a relative change of greater than  $\pm$ 30$\%$ in column density or line intensity.

In this work, some updates produce a relative change of greater than
$\pm$ 90$\%$ in column density. 
 
We find that the two reactions, H$_3$O$^+$ + SiO $\rightarrow$ SiOH$^+$ + H$2$O and
H$_3$+ SiO $\rightarrow$ SiOH$^+$ + H$_2$, produced significant  changes 
in SiOH$^+$ column densities with the updated UDfA2012 rate coefficients. 
The UDfA2012 rate coefficients for the reactions, 
HSiO$_2^+$ + e $\rightarrow$ SiO$_2$ + H and 
OH + SiO $\rightarrow$ SiO$_2$ + H result in convergence problem for high density 
($>$ 10$^{17}$ cm$^{-3}$) and low radiation model. Hence, we retain the UDfA2006 
rate coefficients for these two reactions.   

We calculate the line luminosities/intensities from 
energy levels and
radiative and collisional rate coefficients in the Leiden Atomic and Molecular Database
(LAMDA) format \citep{2005A&A...432..369S}. 
TiO energy levels and radiative rate coefficients are from the Cologne Database for
Molecular Spectroscopy (CDMS) (https://cdms.astro.uni-koeln.de/). The ground state 
of TiO is 
triplet-delta with lambda splitting. We consider 230 energy levels, and the
corresponding 223 radiative transitions. In addition, we consider 444 
collisional transitions with ortho and para H$_2$. The collisional transitions of CO 
with ortho and para H$_2$ are modified from \citet{2010ApJ...718.1062Y} as a proxy 
for ortho and para H$_2$ collisions with TiO. In cases for J$_{low}$=3 there are four possible transitions, so the CO rates were divided by 4. For J$_{low}$=2 and J$_{low}$=1, CO rates were divided by 3 and 2, respectively. Otherwise, each CO rate was divided by 5. The temperature range for collisional 
rates vary from 2 to 3000 K. 
These were compiled and converted to LAMDA format. This will be part of upcoming 2024 release of Cloudy.
In the future, we will replace  
these collisional rates with actual data when they become available. We will also add collisional 
rate coefficients for other colliders, H and He, when data becomes available. 

\subsection{Modelling TiO from VY Canis Majoris}
We present a model for the circumstellar 
envelope of the red supergiant VY CMa to validate our Ti-chemical network. 
All  calculations  
are done using the development version of Cloudy \citep{2023RMxAA..59..327C, 2023RNAAS...7..246G}. VY CMa has a rich inventory 
of various molecules
\citep{2013ApJS..209...38K}. However, here we concentrate mainly on TiO. 

We consider a spherical model with the ionizing source, star, at the centre. Cloudy requires basic input parameters 
such as the radiation field, density at the inner radius which is illuminated by the starlight, chemical abundances, etc. Observations reveal
the effective temperature of the 
central star is 3650 K \citep{2006ApJ...646.1203M} with a luminosity of 
(3$\pm$0.5) $\times$ 10$^5$ L$\odot$ \citep{2008PASJ...60.1007C}. We use this observed value 
to set the SED of the radiation field.

VY CMa is a large star with 
R$\star$ = 1420 $\pm$ 120 R$\odot$ \citep{2012A&A...540L..12W}. We consider the inner radius of the circumstellar envelope to be at R$\star$.
TiO is expected to form nearer the star, where the temperature is high, above the
sublimation temperature of dust. \citet{2016A&A...592A..76D} have observed that NaCl forms 
at R $<$ 250 R$\star$ from the central star. \citet{2013A&A...551A.113K} also  
noticed that 
the main component of the TiO emission covers the same velocity range as the NaCl profile, but 
is confined within R = 28 $\times$ R$\star$. The Dissociation energy of TiO is much higher than NaCl, so in the gas phase it is expected to peak closer to the central star than NaCl. 
Since our main focus is TiO, we further confine our model 
to R = 28 $\times$ R$\star$ from the central star. 

For an oxygen-rich environment, nearly all carbon is locked up in CO. Thus, formation of silicate 
dusts is more favourable than the formation of graphite dust, when physical conditions permit. 
The sublimation temperature of 
silicate dusts is $\approx$ 1400 K. Furthermore, \citet{2013A&A...551A.113K} estimated 
T$_{rot}$ = 1010 $\pm$ 870 K. So, we further constrain our model to have a 
temperature above $>$ 1400 K.

The density, in general, depends on the distance from the central source of ionization. 
We consider the density to be proportional to r$^{-2}$, consistent with a $v\propto r$ wind \citep{{1988ApJ...326..832K},{2006PNAS..10312274Z}}. In addition, it is known that the 
density of the circumstellar envelope can vary over a wide range. Hence, 
in our grid models, 
we vary the density n$_H$(r$_{in}$) (the density at the inner radius (r$_{in}$) of the circumstellar envelope) from 10$^9$ to 10$^{12.5}$ cm$^{-3}$
in steps of 0.25 dex. We consider solar 
abundances of elements as reported in \citet{2010Ap&SS.328..179G}, and we do not 
include dust in our model. Finally, we include a background cosmic ray 
ionization rate (important for the ion balance in molecular regions) of 2 $\times$ 10$^{-16}$ s$^{-1}$  s$^{-1}$ in our grid models \citep{{2008ApJ...675..405S},{2021ApJ...908..138S}}.

\begin{figure}[!t]
\includegraphics[width=\columnwidth]{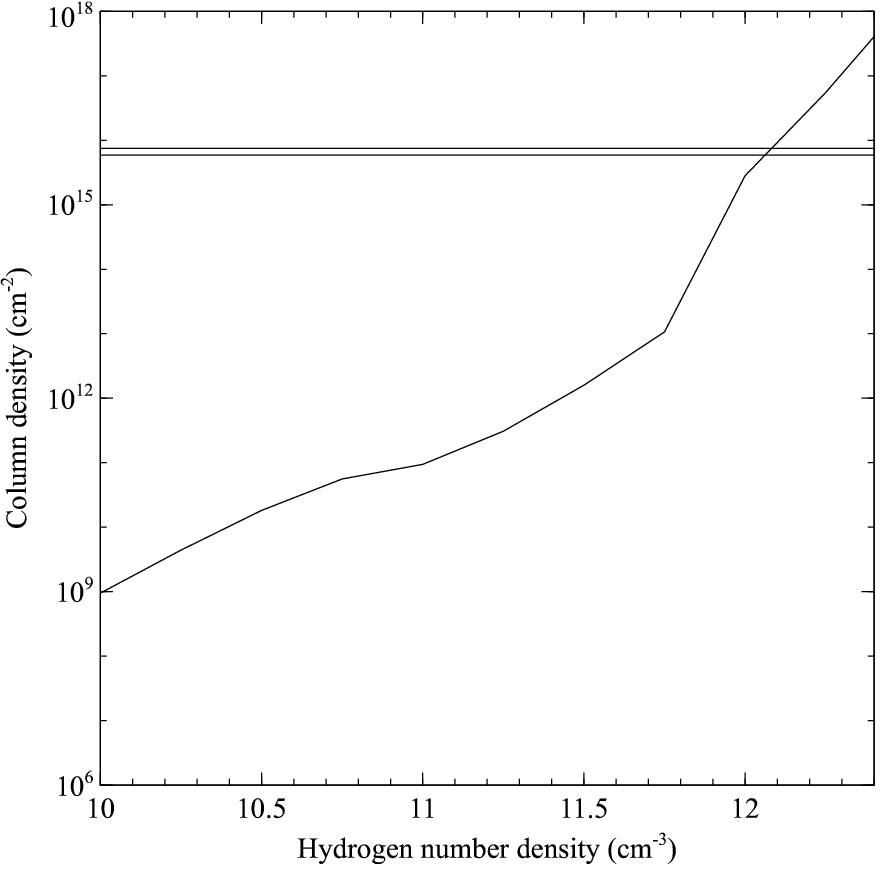}
\caption{Column density of TiO as a function of n$_H$(r$_{in}$). The horizontal lines represent the observed column density range.}
\label{fig:tio_grid_final_1400}
\end{figure}

Figure \ref{fig:tio_grid_final_1400} shows our model's predicted TiO column density
as a function of hydrogen number density, n$_H$(r$_{in}$). The solid horizontal lines represent
the observed column density range \citet{2013A&A...551A.113K}. From the plot, we conclude
that the observed TiO column density comes from a region with n$_H$(r$_{in}$)=
10$^{12.06}$ to 10$^{12.08}$ cm$^{-3}$. 

\begin{figure}[!t]
  \includegraphics[width=\columnwidth]{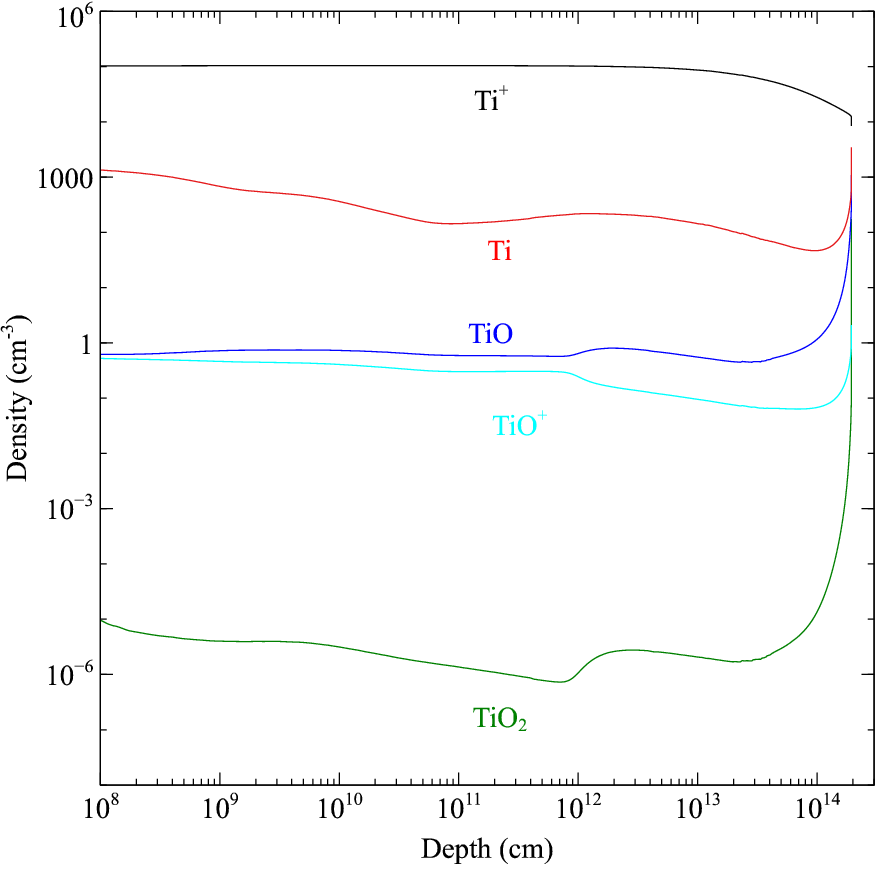}
\caption{Variation of TiO, TiO$^+$, TiO$_2$, Ti, Ti$^+$ as a function of cloud depth. The gas temperature is $>$ 1400 K.}
\label{fig:structure_plot_1400new}
\end{figure}

Next, we consider a single model with n$_H$(r$_{in}$) = 10$^{12.07}$ cm$^{-3}$, 
holding other 
parameters constant. Figure \ref{fig:structure_plot_1400new} shows the variation of TiO, TiO$^+$, 
TiO$_2$, Ti, Ti$^+$ as a function of the depth of the cloud from the illuminated surface. The 
temperature varies from 3431~K to 1400~K across the region considered.
We find that Ti is mostly ionic, and the dominant oxide form of Ti is TiO. 

Our model's predicted neutral hydrogen column density is $\approx$ 8$\times$10$^{25}$ cm$^{-2}$. Although our model does not
include dust, it predicts a high H$_2$ column density $\approx$ 10$^{18}$ cm$^{-2}$. 
This is due to the presence of a 
significant H$^{-}$ abundance which helps to form H$_2$ in dust-free 
environments \citep{2005ApJ...624..794S}. 
However, this single model predicts a 
TiO$_2$ column density, 7.16$\times$10$^{12}$ cm$^{-2}$, much below the observed value. 

\begin{figure}[!t]
\includegraphics[width=\columnwidth]{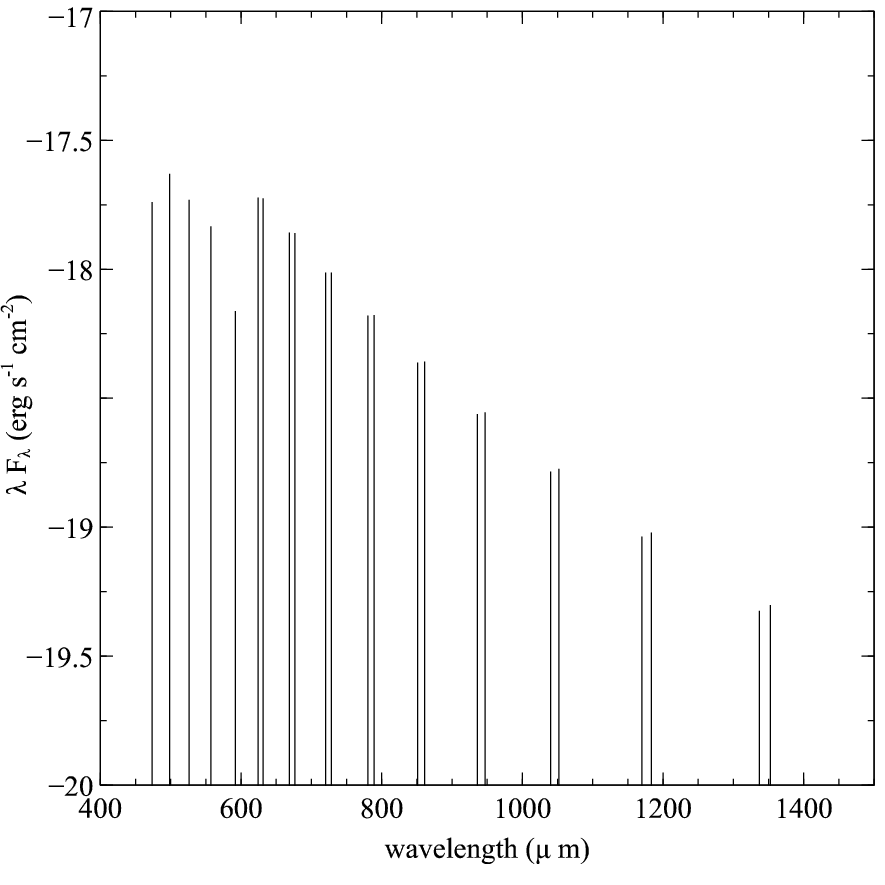}
\caption{Intensity of TiO lines}
\label{fig:intensity_final}
\end{figure}

Figure \ref{fig:intensity_final} shows the line intensity of TiO lines for this single model assuming the 
distance of VY CMa is 1.20 kpc \citep{2012ApJ...744...23Z}.

\begin{figure}[!t]
  \includegraphics[width=\columnwidth]{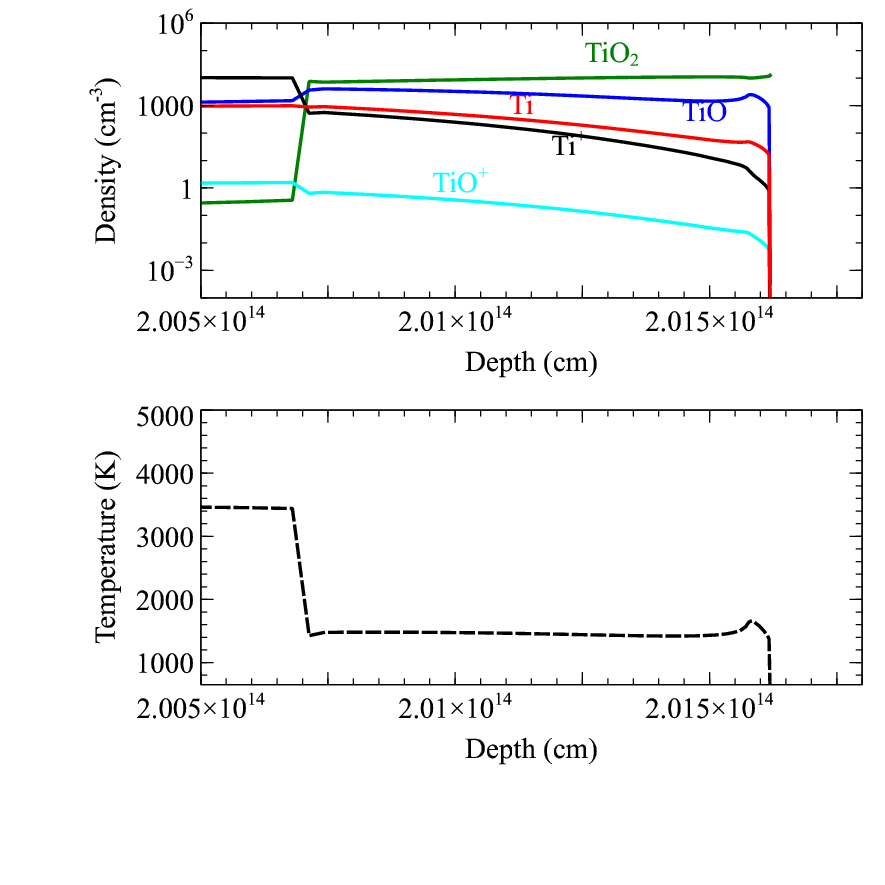}
\caption{Variation of TiO, TiO$^+$, TiO$_2$, Ti, Ti$^+$, and temperature as a function of cloud depth.}
\label{fig:structure_lownew}
\end{figure}

Earlier, 
\citep{1998FaDi..109..303G} have showed that TiO$_2$ forms at a much lower temperature than 
TiO. So, we recomputed the same single model but removing the 1400 K temperature limit. The 
predicted TiO and TiO$_2$ column densities are 7.5$\times$ 10$^{15}$ and 8.7$\times$10$^{15}$ 
cm$^{-2}$, respectively. The observed TiO$_2$ column density is (5.65 $\pm $1.33) $\times$ 
10$^{15}$ cm$^{-2}$ \citep{2015A&A...580A..36D}. Both of these molecules are confined within R = 28 $\times$ R$\star$.
The corresponding ionization structure is shown in Figure \ref{fig:structure_lownew}. 
We notice that as the temperature decreases below 1400 K, more TiO$_2$ is formed. We conclude that
Ti is locked in TiO above 1400 K and in TiO$_2$ at a lower temperature. Since silicate dust
can form below 1400K, we conclude that TiO$_2$ remains in the gas phase outside the dust-formation zone, and might play only a minor role in the dust-condensation process around VY CMa. The same conclusion was made by \citet{2015A&A...580A..36D}. 


\subsection{Effects of input parameters}
There is a debate regarding the luminosity and radius of the star VY CMa. \citet{2006ApJ...646.1203M} have estimated the 
luminosity and radius to be 6$\times$10$^4$ L$\odot$ and 600 R$\odot$, respectively. However, \citet{2012A&A...540L..12W} have
estimated these to be 5$\times$10$^5$ L$\odot$ and R$\star$ = 1420 $\pm$ 120 R$\odot$, respectively. Here, we show the effects of
various input parameters on the predicted hydrogen number density, column densities of TiO and 
TiO$_2$ for the above-mentioned model of VY CMa. 

Firstly, we consider 
a value, 1.0 $\times$ 10$^5$ L$\odot$, instead of 2.5$\times$ 10$^5$ L$\odot$ considered in section 2.2.  
We run a similar grid of models varying n$_H$(r$_{in}$) from 10$^9$ to 10$^{12.5}$ cm$^{-3}$
in steps of 0.25 dex, keeping all the other parameters the same. For this case, the observed TiO column density comes from a 
region with n$_H$(r$_{in}$) =
10$^{11.79}$ to 10$^{11.80}$ cm$^{-3}$. This is lower by $\approx$ 0.3 dex from the value derived in section 2.2. 

Next, we consider a lower value of R$\star$ =770 R$\odot$ compared to the earlier value of R$\star$=1540 R$\odot$ for VY CMa, 
and run a similar grid model keeping other parameters fixed. For this case, the observed TiO column density comes from a 
region with n$_H$(r$_{in}$) = 10$^{12.66}$ to 10$^{12.67}$ cm$^{-3}$. This is higher by $\approx$ 0.6 dex from the value derived in section 2.2.

Further, we change the density to be proportional to r$^{-3}$, and run a similar grid model keeping other parameters fixed as 
described in section 2.2. We find that with the new density law,
the observed TiO column density comes from a 
region with n$_H$(r$_{in}$) = 
10$^{12.28}$ to 10$^{12.31}$ cm$^{-3}$. 
This is higher by $\approx$ 0.2 dex from the value derived in section 2.2. 

We notice that the value of n$_H$(r$_{in}$) is more dependent on the inner radius r$_{in}$ of the envelope. However, for all these cases, as the temperature decreases below 1400 K, more TiO$_2$ is formed. Similarly, both TiO and TiO$_2$ are confined within R = 28 $\times$ R$\star$.

\subsection{Model limitations}
Like every astrophysical model, our model also has some limitations. 
The main limitations are listed below.

Firstly, most of our Ti-related gas-phase chemical reaction rate coefficients involving Ti are copied from analogous Si-related reactions.
Secondly, we have considered only ortho and para--H$_2$ collisional excitation and
deexcitation. 
Thirdly, we have assumed a spherical geometry. However, observations suggest that 
the geometry is not perfectly spherical.
Despite these limitations, our model predicts the observed column density of TiO and TiO$_2$.

\section{Discussions and conclusions}
In this work, we updated gas-phase chemical reaction rates and molecular lines arising
from 
TiO in the spectral synthesis code \textsc{cloudy}. We have added 229 reactions in the chemical 
network. 
In addition, we consider 230 energy levels, corresponding radiative transitions, and 444 
collisional transitions with ortho and para H$_2$. These molecular lines can be observed using ALMA and JWST. 
These updates will be part of the upcoming 2024 release of Cloudy.

Our main conclusions from this work are:

\begin{enumerate}
\item In the gas-phase, Ti is mainly in TiO for temperatures above 1400 K. Whereas, TiO$_2$ dominates at a lower temperature.

\item Our model predicts the observed column densities of TiO and TiO$_2$ around the oxygen rich red super giant VY CMa. Both of these molecules are confined within R = 28 $\times$ R$\star$. However, TiO forms closer to the central star when temperatures are above 1400K.

\item Few ISM-appropriate Ti molecular reaction networks exist due to its low gas-phase abundance in the ISM.
Most of our Ti-related gas-phase chemical reaction rate-coefficients  are copied from analogous Si-related reactions. Our model does predict observed TiO and TiO$_2$ column density around VY CMa. However, we look forward to actual rate coefficients for these reactions. 

\end{enumerate}

In future, we hope to incorporate 
another important precursor of inorganic dust, AlO, into \textsc{Cloudy}'s chemical network.
 
\section{acknowledgments}
We thank Peter van-hoof, Kyle Walker, and Krishal Patel for their valuable help. GS acknowledges support from DST/WOS-A/PM-2/2021.

\bibliography{TiO, ./common/bibliography2} 

\begin{thebibliography}
\expandafter\ifx\csname natexlab\endcsname\relax\def\natexlab#1{#1}\fi
\expandafter\ifx\csname href\endcsname\relax
  \def\href#1#2{}\fi
\expandafter\ifx\csname urllinklabel\endcsname\relax
  \def\urllinklabel{[LINK]}\fi
\expandafter\ifx\csname adsurllinklabel\endcsname\relax
  \def\adsurllinklabel{[ADS]}\fi

\bibitem[{{Chatzikos} {et~al.}(2023){Chatzikos}, {Bianchi}, {Camilloni},
  {Chakraborty}, {Gunasekera}, {Guzm{\'a}n}, {Milby}, {Sarkar}, {Shaw}, {van
  Hoof}, \& {Ferland}}]{2023RMxAA..59..327C}
{Chatzikos}, M., {Bianchi}, S., {Camilloni}, F., {Chakraborty}, P.,
  {Gunasekera}, C.~M., {Guzm{\'a}n}, F., {Milby}, J.~S., {Sarkar}, A., {Shaw},
  G., {van Hoof}, P.~A.~M., \& {Ferland}, G.~J. 2023, \rmxaa, 59, 327


\bibitem[{{Choi} {et~al.}(2008){Choi}, {Hirota}, {Honma}, {Kobayashi},
  {Bushimata}, {Imai}, {Iwadate}, {Jike}, {Kameno}, {Kameya}, {Kamohara},
  {Kan-Ya}, {Kawaguchi}, {Kijima}, {Kim}, {Kuji}, {Kurayama}, {Manabe},
  {Maruyama}, {Matsui}, {Matsumoto}, {Miyaji}, {Nagayama}, {Nakagawa},
  {Nakamura}, {Oh}, {Omodaka}, {Oyama}, {Sakai}, {Sasao}, {Sato}, {Sato},
  {Shibata}, {Tamura}, {Tsushima}, \& {Yamashita}}]{2008PASJ...60.1007C}
{Choi}, Y.~K., {Hirota}, T., {Honma}, M., {Kobayashi}, H., {Bushimata}, T.,
  {Imai}, H., {Iwadate}, K., {Jike}, T., {Kameno}, S., {Kameya}, O.,
  {Kamohara}, R., {Kan-Ya}, Y., {Kawaguchi}, N., {Kijima}, M., {Kim}, M.~K.,
  {Kuji}, S., {Kurayama}, T., {Manabe}, S., {Maruyama}, K., {Matsui}, M.,
  {Matsumoto}, N., {Miyaji}, T., {Nagayama}, T., {Nakagawa}, A., {Nakamura},
  K., {Oh}, C.~S., {Omodaka}, T., {Oyama}, T., {Sakai}, S., {Sasao}, T.,
  {Sato}, K., {Sato}, M., {Shibata}, K.~M., {Tamura}, Y., {Tsushima}, M., \&
  {Yamashita}, K. 2008, \pasj, 60, 1007


\bibitem[{{Churchwell} {et~al.}(1980){Churchwell}, {Hocking}, {Merer}, \&
  {Gerry}}]{1980AJ.....85.1382C}
{Churchwell}, E., {Hocking}, W.~H., {Merer}, A.~J., \& {Gerry}, M.~C.~L. 1980,
  \aj, 85, 1382


\bibitem[{{Dalgarno} \& {McCray}(1973)}]{1973ApJ...181...95D}
{Dalgarno}, A. \& {McCray}, R.~A. 1973, \apj, 181, 95


\bibitem[{{Danilovich} {et~al.}(2020){Danilovich}, {Gottlieb}, {Decin},
  {Richards}, {Lee}, {Kami{\'n}ski}, {Patel}, {Young}, \&
  {Menten}}]{2020ApJ...904..110D}
{Danilovich}, T., {Gottlieb}, C.~A., {Decin}, L., {Richards}, A.~M.~S., {Lee},
  K.~L.~K., {Kami{\'n}ski}, T., {Patel}, N.~A., {Young}, K.~H., \& {Menten},
  K.~M. 2020, \apj, 904, 110


\bibitem[{{De Beck} {et~al.}(2015){De Beck}, {Vlemmings}, {Muller}, {Black},
  {O'Gorman}, {Richards}, {Baudry}, {Maercker}, {Decin}, \&
  {Humphreys}}]{2015A&A...580A..36D}
{De Beck}, E., {Vlemmings}, W., {Muller}, S., {Black}, J.~H., {O'Gorman}, E.,
  {Richards}, A.~M.~S., {Baudry}, A., {Maercker}, M., {Decin}, L., \&
  {Humphreys}, E.~M. 2015, \aap, 580, A36


\bibitem[{{Decin} {et~al.}(2016){Decin}, {Richards}, {Millar}, {Baudry}, {De
  Beck}, {Homan}, {Smith}, {Van de Sande}, \& {Walsh}}]{2016A&A...592A..76D}
{Decin}, L., {Richards}, A.~M.~S., {Millar}, T.~J., {Baudry}, A., {De Beck},
  E., {Homan}, W., {Smith}, N., {Van de Sande}, M., \& {Walsh}, C. 2016, \aap,
  592, A76


\bibitem[{{Ferland} {et~al.}(2017){Ferland}, {Chatzikos}, {Guzm{\'a}n},
  {Lykins}, {van Hoof}, {Williams}, {Abel}, {Badnell}, {Keenan}, {Porter}, \&
  {Stancil}}]{2017RMxAA..53..385F}
{Ferland}, G.~J., {Chatzikos}, M., {Guzm{\'a}n}, F., {Lykins}, M.~L., {van
  Hoof}, P.~A.~M., {Williams}, R.~J.~R., {Abel}, N.~P., {Badnell}, N.~R.,
  {Keenan}, F.~P., {Porter}, R.~L., \& {Stancil}, P.~C. 2017, \rmxaa, 53, 385


\bibitem[{{Ferland} {et~al.}(2013){Ferland}, {Porter}, {van Hoof}, {Williams},
  {Abel}, {Lykins}, {Shaw}, {Henney}, \& {Stancil}}]{2013RMxAA..49..137F}
{Ferland}, G.~J., {Porter}, R.~L., {van Hoof}, P.~A.~M., {Williams}, R.~J.~R.,
  {Abel}, N.~P., {Lykins}, M.~L., {Shaw}, G., {Henney}, W.~J., \& {Stancil},
  P.~C. 2013, \rmxaa, 49, 137


\bibitem[{{Gail} \& {Sedlmayr}(1998)}]{1998FaDi..109..303G}
{Gail}, H.~P. \& {Sedlmayr}, E. 1998, Faraday Discussions, 109, 303


\bibitem[{{Gilli} {et~al.}(2006){Gilli}, {Israelian}, {Ecuvillon}, {Santos}, \&
  {Mayor}}]{2006A&A...449..723G}
{Gilli}, G., {Israelian}, G., {Ecuvillon}, A., {Santos}, N.~C., \& {Mayor}, M.
  2006, \aap, 449, 723


\bibitem[{{Grevesse} {et~al.}(2010){Grevesse}, {Asplund}, {Sauval}, \&
  {Scott}}]{2010Ap&SS.328..179G}
{Grevesse}, N., {Asplund}, M., {Sauval}, A.~J., \& {Scott}, P. 2010, \apss,
  328, 179


\bibitem[{{Gunasekera} {et~al.}(2023){Gunasekera}, {van Hoof}, {Chatzikos}, \&
  {Ferland}}]{2023RNAAS...7..246G}
{Gunasekera}, C.~M., {van Hoof}, P. A.~M., {Chatzikos}, M., \& {Ferland}, G.~J.
  2023, Research Notes of the American Astronomical Society, 7, 246


\bibitem[{{Indriolo} {et~al.}(2007){Indriolo}, {Geballe}, {Oka}, \&
  {McCall}}]{2007ApJ...671.1736I}
{Indriolo}, N., {Geballe}, T.~R., {Oka}, T., \& {McCall}, B.~J. 2007, \apj,
  671, 1736


\bibitem[{{Jorgensen}(1994)}]{1994A&A...284..179J}
{Jorgensen}, U.~G. 1994, \aap, 284, 179


\bibitem[{{Kami{\'n}ski} {et~al.}(2013{\natexlab{a}}){Kami{\'n}ski},
  {Gottlieb}, {Menten}, {Patel}, {Young}, {Br{\"u}nken}, {M{\"u}ller},
  {McCarthy}, {Winters}, \& {Decin}}]{2013A&A...551A.113K}
{Kami{\'n}ski}, T., {Gottlieb}, C.~A., {Menten}, K.~M., {Patel}, N.~A.,
  {Young}, K.~H., {Br{\"u}nken}, S., {M{\"u}ller}, H.~S.~P., {McCarthy}, M.~C.,
  {Winters}, J.~M., \& {Decin}, L. 2013{\natexlab{a}}, \aap, 551, A113


\bibitem[{{Kami{\'n}ski} {et~al.}(2013{\natexlab{b}}){Kami{\'n}ski},
  {Gottlieb}, {Young}, {Menten}, \& {Patel}}]{2013ApJS..209...38K}
{Kami{\'n}ski}, T., {Gottlieb}, C.~A., {Young}, K.~H., {Menten}, K.~M., \&
  {Patel}, N.~A. 2013{\natexlab{b}}, \apjs, 209, 38


\bibitem[{{Keady} {et~al.}(1988){Keady}, {Hall}, \&
  {Ridgway}}]{1988ApJ...326..832K}
{Keady}, J.~J., {Hall}, D. N.~B., \& {Ridgway}, S.~T. 1988, \apj, 326, 832


\bibitem[{Lodders(2002)}]{Lodders_2002}
Lodders, K. 2002, The Astrophysical Journal, 577, 974
 \href{https://dx.doi.org/10.1086/342241}{\urllinklabel}

\bibitem[{{Massey} {et~al.}(2006){Massey}, {Levesque}, \&
  {Plez}}]{2006ApJ...646.1203M}
{Massey}, P., {Levesque}, E.~M., \& {Plez}, B. 2006, \apj, 646, 1203


\bibitem[{{McElroy} {et~al.}(2013){McElroy}, {Walsh}, {Markwick}, {Cordiner},
  {Smith}, \& {Millar}}]{2013A&A...550A..36M}
{McElroy}, D., {Walsh}, C., {Markwick}, A.~J., {Cordiner}, M.~A., {Smith}, K.,
  \& {Millar}, T.~J. 2013, \aap, 550, A36


\bibitem[{{Ram{\'\i}rez} {et~al.}(2020){Ram{\'\i}rez}, {Cridland}, \&
  {Molli{\`e}re}}]{2020A&A...641A..87R}
{Ram{\'\i}rez}, V., {Cridland}, A.~J., \& {Molli{\`e}re}, P. 2020, \aap, 641,
  A87


\bibitem[{{Sch{\"o}ier} {et~al.}(2005){Sch{\"o}ier}, {van der Tak}, {van
  Dishoeck}, \& {Black}}]{2005A&A...432..369S}
{Sch{\"o}ier}, F.~L., {van der Tak}, F.~F.~S., {van Dishoeck}, E.~F., \&
  {Black}, J.~H. 2005, \aap, 432, 369


\bibitem[{{Shaw} {et~al.}(2023){Shaw}, {Ferland}, \&
  {Chatzikos}}]{2023RNAAS...7...45S}
{Shaw}, G., {Ferland}, G., \& {Chatzikos}, M. 2023, Research Notes of the
  American Astronomical Society, 7, 45


\bibitem[{Shaw {et~al.}(2023)Shaw, Ferland, \& Chatzikos}]{Shaw_2023}
Shaw, G., Ferland, G., \& Chatzikos, M. 2023, Research Notes of the AAS, 7, 153
 \href{https://dx.doi.org/10.3847/2515-5172/ace9b5}{\urllinklabel}

\bibitem[{{Shaw} \& {Ferland}(2021)}]{2021ApJ...908..138S}
{Shaw}, G. \& {Ferland}, G.~J. 2021, \apj, 908, 138


\bibitem[{{Shaw} {et~al.}(2005){Shaw}, {Ferland}, {Abel}, {Stancil}, \& {van
  Hoof}}]{2005ApJ...624..794S}
{Shaw}, G., {Ferland}, G.~J., {Abel}, N.~P., {Stancil}, P.~C., \& {van Hoof},
  P.~A.~M. 2005, \apj, 624, 794


\bibitem[{{Shaw} {et~al.}(2022){Shaw}, {Ferland}, \&
  {Chatzikos}}]{2022ApJ...934...53S}
{Shaw}, G., {Ferland}, G.~J., \& {Chatzikos}, M. 2022, \apj, 934, 53


\bibitem[{{Shaw} {et~al.}(2017){Shaw}, {Ferland}, \&
  {Hubeny}}]{2017ApJ...843..149S}
{Shaw}, G., {Ferland}, G.~J., \& {Hubeny}, I. 2017, \apj, 843, 149


\bibitem[{{Shaw} {et~al.}(2020){Shaw}, {Ferland}, \&
  {Ploeckinger}}]{2020RNAAS...4...78S}
{Shaw}, G., {Ferland}, G.~J., \& {Ploeckinger}, S. 2020, Research Notes of the
  American Astronomical Society, 4, 78


\bibitem[{{Shaw} {et~al.}(2008){Shaw}, {Ferland}, {Srianand}, {Abel}, {van
  Hoof}, \& {Stancil}}]{2008ApJ...675..405S}
{Shaw}, G., {Ferland}, G.~J., {Srianand}, R., {Abel}, N.~P., {van Hoof},
  P.~A.~M., \& {Stancil}, P.~C. 2008, \apj, 675, 405


\bibitem[{{Tsai} {et~al.}(2021){Tsai}, {Malik}, {Kitzmann}, {Lyons}, {Fateev},
  {Lee}, \& {Heng}}]{2021ApJ...923..264T}
{Tsai}, S.-M., {Malik}, M., {Kitzmann}, D., {Lyons}, J.~R., {Fateev}, A.,
  {Lee}, E., \& {Heng}, K. 2021, \apj, 923, 264


\bibitem[{{Wakelam} {et~al.}(2012){Wakelam}, {Herbst}, {Loison}, {Smith},
  {Chandrasekaran}, {Pavone}, {Adams}, {Bacchus-Montabonel}, {Bergeat},
  {B{\'e}roff}, {Bierbaum}, {Chabot}, {Dalgarno}, {van Dishoeck}, {Faure},
  {Geppert}, {Gerlich}, {Galli}, {H{\'e}brard}, {Hersant}, {Hickson},
  {Honvault}, {Klippenstein}, {Le Picard}, {Nyman}, {Pernot}, {Schlemmer},
  {Selsis}, {Sims}, {Talbi}, {Tennyson}, {Troe}, {Wester}, \&
  {Wiesenfeld}}]{2012ApJS..199...21W}
{Wakelam}, V., {Herbst}, E., {Loison}, J.~C., {Smith}, I.~W.~M.,
  {Chandrasekaran}, V., {Pavone}, B., {Adams}, N.~G., {Bacchus-Montabonel},
  M.~C., {Bergeat}, A., {B{\'e}roff}, K., {Bierbaum}, V.~M., {Chabot}, M.,
  {Dalgarno}, A., {van Dishoeck}, E.~F., {Faure}, A., {Geppert}, W.~D.,
  {Gerlich}, D., {Galli}, D., {H{\'e}brard}, E., {Hersant}, F., {Hickson},
  K.~M., {Honvault}, P., {Klippenstein}, S.~J., {Le Picard}, S., {Nyman}, G.,
  {Pernot}, P., {Schlemmer}, S., {Selsis}, F., {Sims}, I.~R., {Talbi}, D.,
  {Tennyson}, J., {Troe}, J., {Wester}, R., \& {Wiesenfeld}, L. 2012, \apjs,
  199, 21


\bibitem[{{Welsh} {et~al.}(1997){Welsh}, {Sasseen}, {Craig}, {Jelinsky}, \&
  {Albert}}]{1997ApJS..112..507W}
{Welsh}, B.~Y., {Sasseen}, T., {Craig}, N., {Jelinsky}, S., \& {Albert}, C.~E.
  1997, \apjs, 112, 507


\bibitem[{{Wittkowski} {et~al.}(2012){Wittkowski}, {Hauschildt},
  {Arroyo-Torres}, \& {Marcaide}}]{2012A&A...540L..12W}
{Wittkowski}, M., {Hauschildt}, P.~H., {Arroyo-Torres}, B., \& {Marcaide},
  J.~M. 2012, \aap, 540, L12


\bibitem[{{Yang} {et~al.}(2010){Yang}, {Stancil}, {Balakrishnan}, \&
  {Forrey}}]{2010ApJ...718.1062Y}
{Yang}, B., {Stancil}, P.~C., {Balakrishnan}, N., \& {Forrey}, R.~C. 2010,
  \apj, 718, 1062


\bibitem[{{Zhang} {et~al.}(2012){Zhang}, {Reid}, {Menten}, \&
  {Zheng}}]{2012ApJ...744...23Z}
{Zhang}, B., {Reid}, M.~J., {Menten}, K.~M., \& {Zheng}, X.~W. 2012, \apj, 744,
  23


\bibitem[{{Ziurys}(2006)}]{2006PNAS..10312274Z}
{Ziurys}, L.~M. 2006, Proceedings of the National Academy of Science, 103,
  12274


\end{thebibliography}


\section{appendix}
Here we list the rate coefficients used for various Ti-related reactions in
Table \ref{tab:long}. Ti-related reactions are mainly copied from the analogous 
Si-related reactions from UDfA2006 and UDfA2012. 
For example, in the reference column Si5209 refers a copied reaction from Si 
reaction of UDfA2012 with the reaction number 5209. Ref 1, 2, and 3 refer to \citet{1980AJ.....85.1382C}, \citet{2020A&A...641A..87R}, and \citet{2021ApJ...923..264T}, respectively. The rate coefficient for the reaction, Ti$^+$ + H$^-$ $\rightarrow$ Ti + H, is taken from \citet{1973ApJ...181...95D}.

The format for the two-body rate coefficients are given in an Arrhenius-type equation, 
\begin {equation}
k=\alpha \left(\frac{T}{300}\right)^\beta \, \exp(-\gamma /T).
\end {equation}
where $T$ is the temperature of the gas.
Whereas the photoreaction rate coefficients are given by
\begin{equation}
k= \alpha \, \exp(-\gamma A_{\rm V}) \, .
\end{equation}
where $A_{\rm V}$ is the extinction at visible wavelengths. Details can be found 
in \citet{2013A&A...550A..36M}. For the two-body reactions and photoreactions, $\gamma$ 
depends on the activation energy of the reaction and the increased dust extinction at 
ultraviolet wavelengths, respectively. 
It is to be noted, we do not use UDfA cosmic ray ionization rates. Our cosmic-ray 
ionization rates are normalised
to a total rate for electron production
from cosmic-ray ionization. We adopt a mean cosmic-ray ionization
rate of 2$\times$10$^{-16}$ s$^{-1}$ for atomic hydrogen as the default
background value \citep{{2007ApJ...671.1736I},{2008ApJ...675..405S}}. 
\citet{2021ApJ...908..138S} showed that this rate is dependent of the presence of PAHs as well. 
For detail, see \citet{2022ApJ...934...53S}. 

\clearpage
\onecolumn

\begin{longtable} {|c|c|c|c|c|}
 \caption{List of T\lowercase{i} related chemical reactions \label{tab:long}}\\

 \hline
 \multicolumn{5}{| c |}{Begin of Table}\\
 \hline
 Reactions & $\alpha$ & $\beta$ & $\gamma$ & Ref \\
 \hline
 \endfirsthead

 \hline
 \multicolumn{5}{|c|}{Continuation of Table \ref{tab:long}}\\
 \hline
 Reactions & $\alpha$ & $\beta$ & $\gamma$ & Ref \\
 \hline
 \endhead

\hline
 \multicolumn{5}{|c|}{Continuation of Table \ref{tab:long}}\\
 \hline
 Reactions & $\alpha$ & $\beta$ & $\gamma$ & Ref \\
 \hline
 \endhead

 \hline
 \endfoot

 \hline
 \multicolumn{5}{| c |}{End of Table}\\
 \hline\hline
 \endlastfoot

C + TiH $\rightarrow$ TiC + H&6.59e-11&0&0 &Si5209\\
C + TiH$_2$ $\rightarrow$ HCTi + H&1.e-10&0&0 &Si5207\\
N + TiH $\rightarrow$ TiN + H&1.66e-10&-0.09&0 &Si5513\\
N + TiH$_2$ $\rightarrow$ HNTi + H&8.0e-11&0.17&0 &Si5511\\
N + TiC $\rightarrow$ Ti + CN&5.0e-11&0&0 &Si5509\\
N  + TiC $\rightarrow$ TiN + C&5.e-11&0&0 &Si5510\\
CH$_2$ + Ti $\rightarrow$ HCTi + H&1.e-10&0&0 &Si5241\\
O + TiH $\rightarrow$ TiO + H&1.e-10&0&0 &Si5665\\
O + TiH$_2$ $\rightarrow$ TiO + H$_2$&8.e-11&0&0 &Si5661\\
O + TiH$_2$ $\rightarrow$ TiO + H + H&1.2e-10&0&0 &Si5662\\
O + TiC $\rightarrow$ TiO + C&5.e-11&0&0 &Si5660\\
O + TiC $\rightarrow$ Ti + CO&5.e-11&0&0 &Si5659\\
O + HCTi $\rightarrow$ TiO + CH&2.e-11&0&0 &Si5626\\
O + TiN $\rightarrow$ NO + Ti&5.e-11&0&0 &Si5666\\
O + TiN $\rightarrow$ TiO + N&5.75e-11&0.1&200 &Si5667\\
O + TiNC $\rightarrow$ TiN + CO&1.e-11&0&0 &Si5668\\
O + TiC$_2$ $\rightarrow$ TiC + CO&4.e-11&0&0 &Si5656\\
OH + Ti $\rightarrow$ TiO + H&1.e-10&0&0 &Si5698 \\
OH + TiO $\rightarrow$ TiO$_2$ + H&2.e-12&0&0 &Si5699 \\
C$_2$H$_2$ + Ti $\rightarrow$ TiC$_2$ + H$_2$&1.30e-10&-0.71&29 & Si5100\\
Ti + O$_2$ $\rightarrow$ TiO + O&1.72e-10&-0.53&17 &Si5710\\
H$^+$ + TiH $\rightarrow$ Ti$^+$ + H$_2$&1.70e-9&0&0 & Si2586\\
H + TiH$^+$ $\rightarrow$ Ti$^+$ + H$_2$&1.9e-9&0&0 &Si3069\\
H$^+$ + TiH$_2$ $\rightarrow$ TiH$^+$ + H$_2$&1.5e-9&0&0 & Si2583\\
H$^+$ + HCTi $\rightarrow$ TiC$^+$ + H$_2$&1.5e-9&0&0 & Si2568\\
H$^+$ + HNTi $\rightarrow$ TiN$^+$ + H$_2$&1.5e-9&0&0 & Si2571\\
H + TiS$^+$ $\rightarrow$ HS + Ti$^+$&1.90e-9&0&0 & Si3070\\
H$_2$ + TiC$^+$ $\rightarrow$ HCTi$^+$ + H&1.5e-9&0&0 & Si2693\\
H$_2$ + TiO$^+$ $\rightarrow$ TiOH$^+$ + H&3.2e-10&0&0 & Si2696\\
H$_3^+$ + Ti $\rightarrow$ TiH$^+$ + H$_2$&3.7e-9&0&0 & Si2964\\
H$_3^+$ + TiH $\rightarrow$ TiH$_2^+$ + H$_2$&2.e-9&0&0 & Si2977\\
H$_3^+$ + TiC $\rightarrow$ HCTi$^+$ + H$_2$&3.e-9&0&0 & Si2971\\
H$_3^+$ + TiN $\rightarrow$ HNTi$^+$ + H$_2$&2.e-9&0&0 & Si2978\\
H$_3^+$ + TiO $\rightarrow$ TiOH$^+$ + H$_2$&2.e-9&0&0 & S2981\\
H$_3^+$ + TiS $\rightarrow$ HTiS$^+$ + H$_2$&8.7e-9&-0.5&0 & Si2982\\
He$^+$ + TiH $\rightarrow$ Ti$^+$ + He + H&1.80e-9&0&0 & Si3549\\
He$^+$ + TiH2 $\rightarrow$ Ti$^+$ + He + H$_2$&1.e-9&0&0 & Si3543\\
He$^+$ + TiH2 $\rightarrow$ TiH$^+$ + He + H&1.e-9&0&0 & Si3544\\
He$^+$ + TiC $\rightarrow$ TiC$^+$ + He&2.e-9&-0.5&0 & Si3538\\
He$^+$ + TiC $\rightarrow$ Ti$^+$ + C + He&2.e-9&-0.5&0 & Si3537\\
He$^+$ + HCTi $\rightarrow$ Ti$^+$ + CH + He&1.e-9&0&0 & Si3478\\
He$^+$ + HCTi $\rightarrow$ TiC$^+$ + He + H&1.e-9&0&0 & Si3479\\
He$^+$ + TiN $\rightarrow$ Ti$^+$ + N + He&2.e-9&-0.5&0 & Si3550\\
He$^+$ + HNTi $\rightarrow$ TiN$^+$ + He + H&2.e-9&0&0 & Si3488\\
He$^+$ + TiO $\rightarrow$ Ti$^+$ + O + He&1.e-9&0&0 &1 \\             
He$^+$ + TiO $\rightarrow$ Ti + O$^+$ + He&8.6e-10&-0.5&0 & Si3554\\
He$^+$ + TiC$_2$ $\rightarrow$ Ti$^+$ + C$_2$ + He&2.e-9&-0.5&0 & Si3528\\
He$^+$ + TiNC $\rightarrow$ Ti + CN$^+$ + He&2.e-9&-0.5&0 & Si3551\\
He$^+$ + TiO$_2$ $\rightarrow$ O$_2$ + Ti$^+$ + He&2.e-9&0&0 & Si3552\\
He$^+$ + TiS $\rightarrow$ S + Ti$^+$ + He&3.8e-9&-0.5&0 & Si3556\\
He$^+$ + TiS $\rightarrow$ S$^+$ + Ti + He&3.8e-9&-0.5&0 & Si13555\\
C$^+$ + TiH $\rightarrow$ TiC$^+$ + H&1.1e-9&0&0 & Si1666\\
C + TiH$^+$ $\rightarrow$ TiC$^+$ + H&2.e-10&0&0 & Si2146\\
C$^+$ + TiH$_2$ $\rightarrow$ TiC$^+$ + H2&1.e-9&0&0 & Si1661\\
C$^+$ + TiH$_2$ $\rightarrow$ HCTi$^+$ + H&1.e-9&0&0 & Si1660\\
C + TiH$_2^+$ $\rightarrow$ HCTi$^+$ + H&1.1e-9&0&0 & Si2147\\
C$^+$ + TiC $\rightarrow$ Ti$^+$ + C$_2$&2.50e-9&-0.5&0 & Si1655\\
C$^+$ + HCTi $\rightarrow$ TiC$_2^+$ + H&2.e-9&0&0 & Si1631\\
C$^+$ + TiN $\rightarrow$ TiC$^+$ + N&1.e-9&-0.5&0 & Si1667\\ 
C$^+$ + HNTi $\rightarrow$ TiNC$^+$ + H&2.e-9&0&0 & Si1635\\
C$^+$ + TiO $\rightarrow$ Ti$^+$ + CO&1.e-9&0&0 & 1\\
C + TiO$^+$ $\rightarrow$ Ti$^+$ + CO&1.e-9&0&0 & Si2153\\
C$^+$ + TiS $\rightarrow$ TiC$^+$ + S&2.3e-9&-0.5&0 & Si1669\\
CH + Ti$^+$ $\rightarrow$ TiC$^+$ + H&6.3e-10&-0.5&0 & Si2487\\
CH + TiH$^+$ $\rightarrow$ Ti + CH$_2^+$&6.e-10&-0.5&0 & Si2488\\
CH + TiO$^+$ $\rightarrow$ HCO$^+$ + Ti&5.9e-10&-0.5&0 & Si2489\\
N + TiH$^+$ $\rightarrow$ TiN$^+$ + H&2.e-10&0&0 & Si3692\\
N + TiH$_2^+$ $\rightarrow$ HNTi$^+$ + H&1.e-10&0&0 & Si3693\\
N + TiC$^+$ $\rightarrow$ Ti$^+$ + CN&7.7e-10&0&0 & Si3690\\
N + TiO$^+$ $\rightarrow$ NO$^+$ + Ti&9.e-11&0&0 & Si3696\\
N + TiO$^+$ $\rightarrow$ NO + Ti$^+$&2.1e-10&0&0 & Si3697\\
CH$_2$ + Ti$^+$ $\rightarrow$ HCTi$^+$ + H&8.7e-10&0&0 & Si2239\\
CH$_2$ + TiO$^+$ $\rightarrow$ H$_2$CO + Ti$^+$&7.88e-10&0&0 & Si1589\\
NH + Ti$^+$ $\rightarrow$ TiN$^+$ + H&1.e-9&-0.5&0 & Si3846\\
CH$_3^+$ + Ti $\rightarrow$ HCTi$^+$ + H$_2$&2.e-10&0&0 & Si2318\\
CH$_3$ + Ti$^+$ $\rightarrow$ HCTi$^+$ + H$_2$&1.e-9&0&0 & Si2323\\
O + TiH$^+$ $\rightarrow$ TiO$^+$ + H&4.e-10&0&0 & Si3977\\
O + TiH$_2$+ $\rightarrow$ TiOH$^+$ + H&6.3e-10&0&0 & Si3978\\
O + TiC$^+$ $\rightarrow$ TiO$^+$,C&6.e-10&0&0 & Si3974\\
O + TiO$^+$ $\rightarrow$ O$_2$ + Ti$^+$&2.e-10&0&0 & Si3985\\
O + TiNC$^+$ $\rightarrow$ TiN$^+$ + CO&1.e-10&0&0 & Si3983\\
OH + Ti$^+$ $\rightarrow$ TiO$^+$ + H&6.3e-10&-0.5&0 & Si4030\\
OH$^+$ + Ti $\rightarrow$ TiH$^+$ + O&1.9e-9&0&0 & Si4012\\
OH$^+$ + TiH $\rightarrow$ TiH$_2^+$ + O&1.e-9&0&0 & Si4014\\
OH$^+$ + TiC $\rightarrow$ HCTi$^+$ + O&4.5e-9&-0.5&0 & Si4013\\
OH$^+$ + TiO $\rightarrow$ TiOH$^+$ + O&9.e-10&-0.5&0 & Si4015\\
NH$_3$ + TiH$^+$ $\rightarrow$ Ti + NH$_4^+$&1.e-9&-0.5&0 & Si3824\\
NH$_3$ + TiOH$^+$ $\rightarrow$ NH$_4^+$ + TiO&2.5e-9&-0.5&0 & Si3825\\
NH$_3$ + HTiS$^+$ $\rightarrow$ NH$_4^+$ + TiS&9.7e-10&-0.5&0 & Si3812\\
H$_2$O + Ti$^+$ $\rightarrow$ TiOH$^+$ + H&2.30e-10&-0.5&0 & Si2790\\
H$_2$O + TiH$^+$ $\rightarrow$ Ti + H$_3$O$^+$&8.e-10&-0.5&0 & Si2792\\
H$_2$O + HNTi$^+$ $\rightarrow$ TiN + H$_3$O$^+$&2.e-9&-0.5&0 & Si2772\\
H$_2$O + HTiS$^+$ $\rightarrow$ TiOH$^+$ + H$_2$S&1.1e-9&-0.5&0 & Si2777\\
H$_3$O$^+$ + Ti $\rightarrow$ TiH$^+$ + H$_2$O&1.8e-9&0&0 & Si3039\\
H$_3$O$^+$ + TiH $\rightarrow$ TiH$_2^+$ + H$_2$O&9.7e-10&0&0 & Si3045\\
H$_3$O$^+$ + TiC $\rightarrow$ HCTi$^+$ + H$_2$O&3.e-9&-0.5&0 & Si3042\\
H$_3$O$^+$ + TiO $\rightarrow$ TiOH$^+$ + H$_2$O&2.0e-9&-0.5&0 & Si3047\\
HF + Ti$^+$ $\rightarrow$ TiF$^+$ + H&5.7e-9&-0.5&0 & Si3236\\
C$_2$ + TiO$^+$ $\rightarrow$ TiC$^+$ + CO&7.6e-10&0&0 & Si1686\\
C$_2$H + Ti$^+$ $\rightarrow$ TiC$_2^+$ + H&1.e-9&0&0 & Si1972\\
C$_2$H$_2^+$ + Ti $\rightarrow$ TiC$_2^+$ + H$_2$&2.e-10&0&0 & Si1750\\
HCN + Ti$^+$ $\rightarrow$ TiNC$^+$ + H&1.4e-12&-0.5&0 & Si3106\\
HNC + Ti$^+$ $\rightarrow$ TiNC$^+$,H&1.e-9&-0.5&0 & Si3267\\
HCN + HTiS$^+$ $\rightarrow$ HCNH$^+$ + TiS&6.1e-10&-0.5&0 & Si3101\\
CO + HNTi$^+$ $\rightarrow$ TiN + HCO$^+$&2.e-9&0&0 & Si2508\\
CO + TiO$^+$ $\rightarrow$ CO$_2$ + Ti$^+$&7.9e-10&0&0 & Si2513\\
HCNH$^+$ + Ti $\rightarrow$ TiNC$^+$ + H$_2$&5.e-10&0&0 & Si3117\\
H$_2$NC$^+$ + Ti $\rightarrow$ TiNC$^+$ + H$_2$&5.e-10&0&0 & Si2723\\
Ti + HCO$^+$ $\rightarrow$ TiH$^+$ + CO&3.3e-10&0&0 & S3210\\
Ti$^+$ + C$_2$H$_5$OH $\rightarrow$ TiOH$^+$ + C$_2$H$_5$&2.2e-9&-0.5&0 & Si4080\\
Ti$^+$ + HC$_4$H $\rightarrow$ C$_4$H$^+$ + TiH&1.6e-9&0&0 & SiRATE12\\
Ti$^+$ + NCCN $\rightarrow$ TiNC$^+$ + CN&1.5e-10&0&0 & Si4091\\
Ti$^+$ + OCS $\rightarrow$ TiS$^+$ + CO&9.e-10&0&0 & Si409\\ 
HCO$^+$ + TiH $\rightarrow$ TiH$_2^+$ + CO&8.7e-10&0&0 & Si3223\\
HCO$^+$ + TiC $\rightarrow$ HCTi$^+$ + CO&2.e-9&-0.5&0 & Si3218\\
HCO$^+$ + TiO $\rightarrow$ TiOH$^+$ + CO&7.9e-10&-0.5&0 & Si3226\\
HCO$^+$ + TiS $\rightarrow$ HTiS$^+$ + CO&3.3e-9&-0.5&0 & Si3227\\
TiH + S$^+$ $\rightarrow$ TiS$^+$ + H&4.2e-10&0&0 & Si4106\\
TiH$_2^+$ + O$_2$ $\rightarrow$ TiOH$^+$ + OH&2.4e-11&0&0 & Si4103\\
TiH$_2^+$ + S $\rightarrow$ HTiS$^+$ + H&1.1e-9&0&0 & Si4104\\
O$_2$ + TiS$^+$ $\rightarrow$ SO$^+$ + TiO&6.23e-11&0&0 & Si3896\\
O$_2$ + TiS$^+$ $\rightarrow$ TiO$^+$ + SO&2.67e-11&0&0 & Si3897\\
S + TiO$^+$ $\rightarrow$ SO + Ti$^+$&1.e-9&0&0 & Si4063\\
H$_2$S + HTiS$^+$ $\rightarrow$ H$_3$S$^+$ + TiS&2.9e-10&-0.5&0 & Si2801\\
H$^+$ + Ti $\rightarrow$ Ti$^+$ + H&9.9e-10&0&0 & Si418 \\
H$^+$ + TiH $\rightarrow$ TiH$^+$ + H&1.7e-9&0&0 & Si431\\
H$^+$ + TiH$_2$ $\rightarrow$ TiH$_2^+$ + H&1.5e-9&0&0 & Si428\\
H$^+$ + TiC $\rightarrow$ TiC$^+$ + H&3.e-9&-0.5&0 & Si425\\
H$^+$ + HCTi $\rightarrow$ HCTi$^+$ + H&1.5e-9&0&0 & Si392\\
H$^+$ + TiN $\rightarrow$ TiN$^+$ + H&3.e-9&-0.5&0 & Si432\\
H$^+$ + HNTi $\rightarrow$ HNTi$^+$ + H&1.5e-9&0&0 & Si394\\
H$^+$ + TiO $\rightarrow$ TiO$^+$ + H&1.e-9&0&0 & 1\\
H$^+$ + TiC$_2$ $\rightarrow$ TiC$_2^+$ + H&3.e-9&-0.5&0 & Si419\\
H$^+$ + TiNC $\rightarrow$ TiNC$^+$ + H&3.e-9&-0.5&0 & Si433\\
H$^+$ + TiS $\rightarrow$ TiS$^+$ + H&1.50e-9&-0.5&0 & Si435\\
He$^+$ + Ti $\rightarrow$ Ti$^+$ + He&3.3e-9&0&0 & Si528\\
C$^+$ + Ti $\rightarrow$ Ti$^+$ + C&2.1-9&0&0 & Si194\\
C$^+$ + TiH$_2$ $\rightarrow$ TiH$_2^+$ + C&1.e-9&0&0 & Si201\\
C$^+$ + TiC $\rightarrow$ TiC$^+$  + C&2.50e-9&-0.5&0 & Si198\\
C$^+$ + TiN $\rightarrow$ TiN$^+$ + C&1.e-9&-0.5&0 & Si203\\
C$^+$ + TiC$_2$ $\rightarrow$ TiC$_2^+$ + C&2.e-9&-0.5&0 & Si195\\
C$^+$ + TiS $\rightarrow$ TiS$^+$ + C&2.3e-9&-0.5&0 & Si204\\
CH$^+$ + Ti $\rightarrow$ Ti$^+$ + CH&2.e-10&0&0 & Si257\\
NH$_3^+$ + Ti $\rightarrow$ Ti$^+$ + NH$_3$&1.9e-9&0&0 & Si597\\
H$_2$O$^+$ + Ti $\rightarrow$ Ti$^+$ + H$_2$O&3.e-9&0&0 & Si477\\
Na + Ti$^+$ $\rightarrow$ Ti + Na$^+$&2.7e-9&0&0 & Si645\\
Mg + TiO$^+$  $\rightarrow$ TiO + Mg$^+$&1.e-9&0&0 & Si541\\
Ti + NO$^+$ $\rightarrow$ NO + Ti$^+$&1.6e-9&0&0 & Si720\\
Ti + P$^+$ $\rightarrow$ P + Ti$^+$&1.e-9&0&0 &Si722\\
Ti + O$_2^+$ $\rightarrow$ O$_2$ + Ti$^+$&1.6e-9&0&0 & Si721\\
Ti + S$^+$ $\rightarrow$ S + Ti$^+$&1.6e-9&0&0 & Si723\\
Ti + HS$^+$ $\rightarrow$ HS + Ti$^+$&1.4e-9&0&0 & Si719\\
Ti + H$_2$S$^+$ $\rightarrow$ H$_2$S + Ti$^+$&1.60e-9&0&0 & Si718\\
Ti + CS$^+$ $\rightarrow$ CS + Ti$^+$&1.50e-10&0&0 & Si716\\
HCO + TiO$^+$ $\rightarrow$ TiO + HCO$^+$&6.6e-10&-0.5&0 & Si504\\
TiH + S$^+$ $\rightarrow$ S + TiH$^+$&4.2e-10&0&0 &Si724\\
NO + TiO$^+$ $\rightarrow$ TiO + NO$^+$&7.2e-10&0&0 & Si631\\
S$^+$ + TiC $\rightarrow$ TiC$^+$ + S&3.7e-9&-0.5&0 & Si705\\
S$^+$ + TiS $\rightarrow$ TiS$^+$ + S&3.2e-9&-0.5&0 & Si706\\
TiO$^+$ + Fe $\rightarrow$ Fe$^+$ + TiO&1.e-9&0&0 & Si725\\
C$^-$ + Ti$^+$ $\rightarrow$ Ti + C&7.51e-8&-0.5&0 & Si4138\\
Ti$^+$ + S$^-$ $\rightarrow$ S + Ti&7.51e-8&-0.5&0 & Si5084\\
TiH$^+$ + e$^-$ $\rightarrow$ Ti + H&2.e-7&-0.5&0 & Si1498\\
TiH$_2^+$ + e$^-$ $\rightarrow$ Ti + H + H&2.e-7&-0.5&0 & Si1500\\
TiH$_2^+$ + e$^-$ $\rightarrow$ Ti + H$_2$&1.5e-7&-0.5&0 & Si1499\\
TiH$_2^+$ + e$^-$ $\rightarrow$ TiH + H&1.5e-7&-0.5&0 & Si1501\\
TiC$^+$ + e$^-$ $\rightarrow$ Ti + C&2.e-7&-0.5&0 & Si1471\\
HCTi$^+$ + e$^-$ $\rightarrow$ Ti + CH&1.5e-7&-0.5&0 & Si1362\\
HCTi$^+$ + e$^-$ $\rightarrow$ TiC + H&1.5e-7&-0.5&0 & Si1363\\
TiN$^+$ + e$^-$ $\rightarrow$ Ti + N&2.0e-7&-0.5&0 & Si1508 \\ 
HNTi$^+$ + e$^-$ $\rightarrow$ Ti + NH&1.5e-7&-0.5&0 & Si1376\\
HNTi$^+$ + e$^-$ $\rightarrow$ TiN + H&1.5e-7&-0.5&0 & Si1377\\
TiO$^+$ + e$^-$ $\rightarrow$ Ti + O&1.e-7&0&0 & 1\\       
TiOH$^+$ + e$^-$ $\rightarrow$ Ti + OH&1.5e-7&-0.5&0 & Si1515\\
TiOH$^+$ + e$^-$ $\rightarrow$ TiO + H&1.5e-7&-0.5&0 & Si1516\\
TiF$^+$ + e$^-$ $\rightarrow$ Ti + F&2.0e-7&-0.5&0 & Si1497  \\ 
TiC$_2^+$ + e$^-$ $\rightarrow$ Ti + C$_2$&1.5e-7&-0.5&0 & Si1472\\
TiC$_2$+ + e$^-$ $\rightarrow$ TiC + C&1.5e-7&-0.5&0 & Si1473\\
TiNC$^+$ + e$^-$ $\rightarrow$ Ti + CN&3.0e-7&-0.5&0 & Si1509\\
TiS$^+$ + e$^-$ $\rightarrow$ S + Ti&2.0e-7&-0.5&0 & Si1517\\
HTiS$^+$ + e$^-$ $\rightarrow$ HS + Ti&1.5e-7&-0.5&0 &Si1400\\
HTiS$^+$ + e$^-$ $\rightarrow$ TiS + H&1.5e-7&-0.5&0 & Si1401\\
Ti$^+$ + e$^-$ $\rightarrow$ Ti + PHOTON&4.26e-12&-0.62&0 & Si6178\\
H + Ti$^+$ $\rightarrow$ TiH$^+$ + PHOTON&1.17e-17&-0.14&0 & Si6127\\
H$_2$ + Ti$^+$ $\rightarrow$ TiH$_2^+$ + PHOTON&3.e-18&0&0 & Si6114\\
O + Ti$^+$ $\rightarrow$ TiO$^+$ + PHOTON&9.22e-19&-0.08&-21.2 & Si6136\\
O + Ti $\rightarrow$ TiO + PHOTON&5.52e-18&0.31& & Si6137\\
Ti + PHOTON $\rightarrow$ Ti$^+$ + e-&3.1e-9&0&2.3 & Si6017\\
TiH + PHOTON $\rightarrow$ Ti + H&2.8e-9&0&1.6 & Si6038\\
TiH$^+$ + PHOTON $\rightarrow$ Ti$^+$ + H&2.7e-9&0&1.2 & Si6029\\
TiC + PHOTON $\rightarrow$ Ti + C&1.0e-10&0&2.3 & Si6025\\
TiO + PHOTON $\rightarrow$ Ti + O&1.6e-9&0&2.3 & Si6044\\
TiO + PHOTON $\rightarrow$ TiO$^+$ + e$^-$&2.4e-10&0&2 & Si6045\\
TiO$^+$ + PHOTON $\rightarrow$ Ti$^+$ + O&1.0e-10&0&2 & Si6042\\
TiS + PHOTON $\rightarrow$ S + Ti&1.0e-10&0&2.3 & Si6046\\
Ti + CRPHOT $\rightarrow$ Ti$^+$ + e$^-$&2115&0&0 & Si4424\\
TiH + CRPHOT $\rightarrow$ Ti + H&250&0&0 &Si980\\
TiC + CRPHOT $\rightarrow$ Ti + C&250&0&0 & Si974\\
TiO + CRPHOT $\rightarrow$ Ti + O&250&0&0 & Si984\\
TiS + CRPHOT $\rightarrow$ S + Ti&250&0&0 & Si985\\
Ti + O $\rightarrow$ TiO$^+$ + e$^-$&1.0e-11&0&1973 & chu80,opp77\\
TiO$^+$ + e$^-$ $\rightarrow$ TiO + PHOTON&5.0e-10&0&0 & chu80\\
Ti$^+$ + O$_2$ $\rightarrow$ TiO$^+$ + O&1.0e-9&0&0 & 1\\
Ti$^+$ + H$_2$O $\rightarrow$ TiO$^+$ + H$_2$&1.0e-9&0&0 & 1\\
TiO$^+$ + Ca $\rightarrow$ TiO + Ca$^+$&1.0e-9&0&0 & chu80\\
TiO$^+$ + Na $\rightarrow$ TiO + Na$^+$&1.0e-9&0&0 & chu80\\
TiO$^+$ + Al $\rightarrow$ TiO + Al$^+$&1.0e-9&0&0 & chu80\\
TiO$^+$ + K $\rightarrow$ TiO + K$^+$&1.0e-9&0&0 & chu80\\
TiO + C $\rightarrow$ Ti + CO&1.0e-10&0&0 & chu80\\
Ti$^+$ + CH$_3$OH $\rightarrow$ TiOH$^+$ + CH$_3$&1.65e-9&-0.5&0 & Si4088 \\
TiH + H $\rightarrow$ Ti + H$_2$&8.3e-11&0&0 & 2\\
TiH + O $\rightarrow$ Ti + OH&1.66e-10&0&0 & 2\\
TiH + CH$_3$ $\rightarrow$ Ti + CH$_4$&1.0e-10&0&0 & 2\\
TiC + H $\rightarrow$ Ti + CH&1.0e-10&0&20109 &2\\
Ti + CO $\rightarrow$ TiC + O&8.0e-10&0&68842 & 2\\
Ti + CN $\rightarrow$ TiC + N&5.0e-10&0&29543 & 2\\     
Ti + NO $\rightarrow$ TiN + O&5.0e-11&0&19706. & 2\\ 
TiN + H $\rightarrow$ Ti + NH&1.e-10&0&15655 & 2\\
Ti + NO $\rightarrow$ TiO + N&9.e-11&-0.96&0 &  2 \\
TiO$_2$ + O $\rightarrow$ TiO + O$_2$&1.e-10&0&11877 & 2\\
TiO$_2$ + CO $\rightarrow$ TiO + CO$_2$&1.e-10&0&9160 & 2\\
TiO + O $\rightarrow$ Ti + O$_2$&5.42e-13&0&20794 & 2\\
TiO + N $\rightarrow$ Ti + NO&4.76e-12&0&4772 & 2\\
TiO + N$_2$ $\rightarrow$ Ti + N$_2$O&5.9e-12&0&62193 & 2\\
Ti + CO$_2$ $\rightarrow$ TiO + CO&7.01e-11&0&1790 & 2 \\   
TiO + NO $\rightarrow$ TiO$_2$ + N&7.8e-12&0&0 & 3\\  
TiO + O$_2$ $\rightarrow$ TiO$_2$ + O&1.5e-12&0&0 & 3\\
TiO + N$_2$O $\rightarrow$ TiO$_2$ + N$_2$ & 4.e-13&0&0 & 3\\
\end{longtable}
\twocolumn

\end{document}